# INTEGRATING OPC DATA INTO GSN INFRASTRUCTURES

Olivier PASSALACQUA, Eric BENOIT, Marc-Philippe HUGET, Patrice MOREAUX
*Laboratoire d'Informatique, Systèmes, Traitement de l'Information et de la Connaissance*
*Universitée de Savoie, B.P. 80439, 74944 Annecy le Vieux Cedex, France*
*{olivier.passalacqua, eric.benoit, marc-philippe.huget, patrice.moreaux}@univ-savoie.fr*

**ABSTRACT**

This paper presents the design and the implementation of an interface software component between OLE for Process Control (OPC) formatted data and the Global Sensor Network (GSN) framework for management of data from sensors. This interface, named wrapper in the GSN context, communicates in Data Access mode with an OPC server and converts the received data to the internal GSN format, according to several temporal modes. This work is realized in the context of a Ph.D. Thesis about the control of distributed information fusion systems. The developed component allows the injection of OPC data, like measurements or industrial processes states information, into a distributed information fusion system deployed in a GSN framework. The component behaves as a client of the OPC server. Developed in Java and based on the Opensaca Utgard, it can be deployed on any computation node supporting a Java virtual machine. The experiments show the component conformity according to the Data Access 2.05a specification of the OPC standard and to the temporal modes.The paper must have an abstract.

## 1. INTRODUCTION

Last decade has seen more and more applications based on information fusion. These applications range from military domain, such as target detection, to everyday life with house equipments. Moreover industrial workshops use an increasing number of various sensors for machine control as well as for information processes. It is then interesting to allow these information systems to benefit from information fusion capabilities. In this perspective, we have to interconnect industrial data systems with Information Fusion Systems (IFS).

On one side the area of industrial data systems, where the OLE for Process Control standard (OPC)[OPC] is largely deployed since the medal of the nineties and this standard is continuously evolving. Even if the OPC foundation provides a large panel of specifications, the most used is the Data Access one, which allows a basic information share. OPC allows interconnection of data sources and data sink through communication managers called OPC *servers*. Data sources and sinks termed as OPC *clients* may be industrial equipments, measure devices, alarm buttons, etc. Although last version of the OPC standard introduces communication based on Intranet and Internet protocols, a large part of installed OPC networks uses the COM/DCOM interaction model developed in Microsoft Windows environments.

On the other side, there exist a lot of IFS, either academic or commercial. They address different kinds of data sources ranging from low power sensors to huge databases. Some systems are very specialized, some are able to be deployed on many kinds of processing nodes. In the perspective of our research activities, we are using the Global Sensor Network (GSN)[Sal07] as an experimental IFS. GSN allows distributed and dynamic IFS deployment on several computers equipped with several input converters (termed "wrappers"). However, there exist no wrapper for connecting OPC data sources to GSN.

Several approaches can be used to connect OPC servers to GSN. Since we will mostly deal with installed OPC systems, we have chosen not to impact the OPC configuration. Consequently, our wrapper behaves exactly as an OPC client getting data from one OPC server in a COM-DCOM context. It then formats these data in the GSN data format (termed StreamElement in the GSN context).

This paper presents the main features and the software architecture of the GSN system in Section 2. Section 3 recalls the OPC standard. The design and the implementation of our wrapper is exposed in Section 4. An example of error detection is given as a possible application in Section 5. Section 6 summarizes our work and proposes future works.

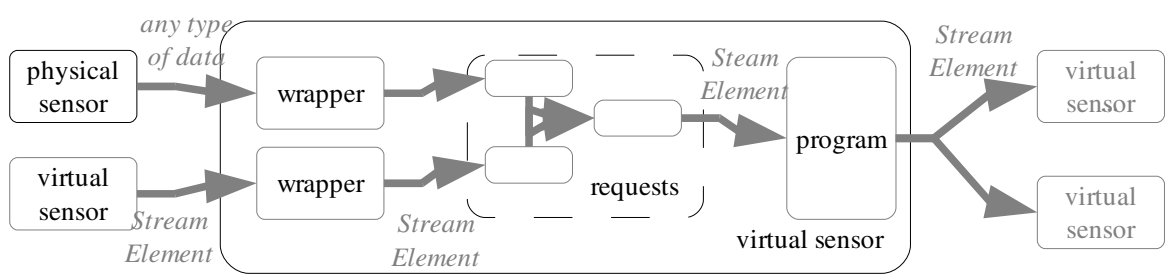

*Figure 1: Virtual sensor structure*

## 2. GSN

The Global Sensor Network project (GSN) [AHS07a] is a middleware supported by research teams in EPFL (Lausanne, Switzerland) and DERI (Galway, Ireland). Its purpose is to get data from various sources such as sensor networks, and to process them through functional nodes. A general GSN system is made of several nodes called GSN cores (nodes) and sensor networks connected to these GSN cores. Several GSN cores may be interconnected through an underlying network of computers. Each core has a default data-base system used to store and to process information. A detailed presentation of GSN is given in [Sal07].

**Wrapper, virtual sensor and GSN core.** A GSN core manages a set of Virtual Sensors (VSs), usually partially connected. VSs are the main elements of the GSN information process. A VS (see Figure 1) selects information received by its sources, termed {*wrappers* in GSN, aggregates some of these selected information, optionally applies a specific processing to this aggregate and finally transmits the result to another VS or to an entity external to GSN. Wrappers are the links between information sources outside the VS and the VS. Sources may be sensors networks (outside of GSN) or output of other VSs. A wrapper is in charge of the communication with its sources. It formats its output data in GSN compatible data structures termed *StreamElements*. Information selection from each wrapper is carried out by a SQL-like query (we call it a *wrapper request*) and result is stored into a data-base. Following information aggregation is then processed also with SQL-like queries (let us call it the *global request*) on these results possibly from several wrapper output selections. Possible specific final processing and result output is achieved by an instance of a Java class defined by the designer of the VS.

**StreamElement** are data units processed by GSN. Such an element is first composed of an array of strings with the name, the type and the description of the produced elements. Then there is a second array with the value of the produced elements and at last there is a timestamp.

Let us emphasize that the resulting data flow in GSN is a flow of (discrete time) data elements (the StreamElements) even if the input of a wrapper may be continuous data.

User's configuration. The configuration of a VS network is defined by the link between the VS and the wrapper they use. The GSN user can connect a new type of information source by developing a new Java class according to the GSN specifications. The new class is then loaded at the GSN core start up before the execution of the fusion process.

**VS description.** At startup, a GSN core read the configurations of its VSs from XML description files (one for each VS), and loads required Java code into its JVM. A VS description file is divided into three parts and describes the architecture of the VS, which requests should apply, what types of data are processed and what is the Java class of the VS. It also defines symbolic names for the various data flows managed by the VS and its wrappers.

**Dynamic configuration**. A GSN core is continuously aware of all XML VS description files in a specific directory and it retrieves at startup the list of its possible wrappers in a file with couples (GSN wrapper name, wrapper Java class name). Java classes of the wrappers has to be available at startup. Thus GSN cannot, at this time of writing, handle dynamic injection of new wrappers. However, a GSN core user can force it to start and stop a VS simply by putting/removing its XML description file in the dedicated directory.

## 3. OPC

Object linking and embedding for Process Control (OPC) is a set of specifications initially written by the WinSEM (Windows for Science Engineering and Manufacturing) group to standardize interfaces between devices and applications in the context of workshops and Microsoft Windows Operating Systems. Since 1996 new OPC specifications are released and updated. Initially, data exchanges between OPC servers and clients were based on the COM-DCOM standard [Roc98]. However, interoperability with more and more various devices, not necessary running on a Microsoft Windows environment, leads the OPC foundation to define a new general OPC context and new communication protocols [Zhe02]. This last OPC specification [LM06], named Unified Architecture, is using a unique Web Service interface and all operations such as Data Access, Alarm and Even and others (see [BHMP07] for details) have been redefined in this context.

Created more than 10 years ago, OPC protocol is still omnipresent in the factories and the industrial automata sellers mainly produce machines only compatible with the Data Access specification.

**Data-access specification.** The first introduced specification was the **Data Access** (DA) specification which is a request/response mode. In this mode, the OPC server can be seen as a shared memory where the OPC clients can read or write values, and the role of the server is to manage concurrent accesses to data. This is the most used communication mode in installed OPC systems and it is well adapted to measurement extraction from a workshop for processing outside the workshop (all communication modes are in [OPC]). Our OPC-GSN wrapper conforms to the DA specification as the shown in its technical report [BHMP07]. The other specifications are not retained because they define the OPC server not as an information source or because they introduce redundant mechanisms or GSN incompatible data as also discussed in [BHMP07].

## 4. OPC INTEGRATION IN GSN

The goal of the OPC-GSN wrapper is to allow a GSN system to get data from an OPC system. In this first version, we consider the OPC system as a data source only and we do not control the OPC system. Information fusion functions are provided by the GSN system and the OPC server is communicating with the wrapper in the DA mode. Since OPC servers expose data items at any time and GSN infrastructures deal with discrete data flows, the first design problem is to define a timed conversion semantics between OPC data and GSN StreamElement flows. The second main design choice is related to the position of the converter in the software/hardware architecture made of the OPC server and its connected GSN core.

### 4.1 Timed semantics of data transfer from OPC to GSN

Discrete data flows in IFS such as GSN are sequences of produced and consumed information elements. Here the StreamElements are produced by wrappers and consumed by VSs. On one hand the OPC server usually updates its item values with a given sampling period $\Delta t_S$, and on the other hand the converter should request the OPC server every update periods $\Delta t_U$.

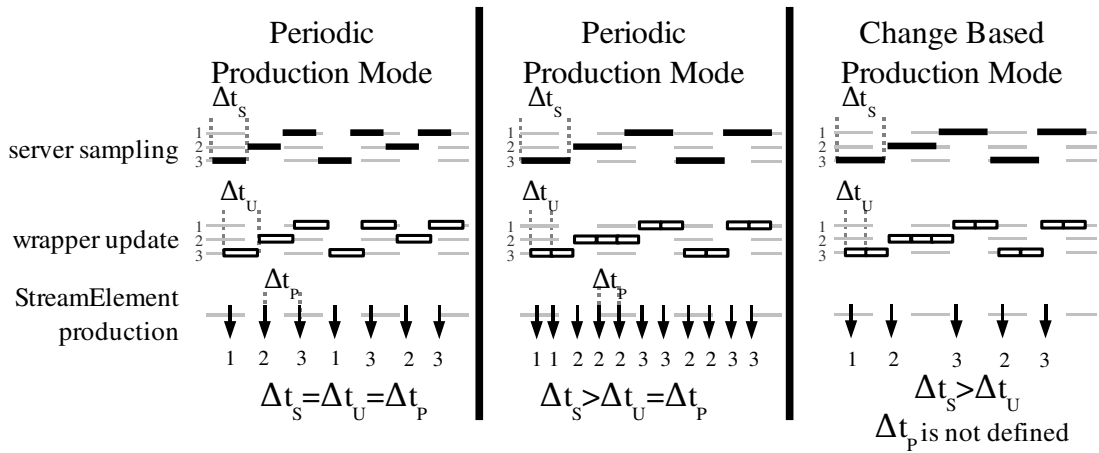

Figure 2: Conversion modes from OPC values to GSN StreamElements

For what concerns the converter, at least two production modes can be defined. Either the converter produces a StreamElement at each update time and the production period $\Delta t_P$ is the same as the update period $\Delta t_U$ - this is the Periodic Production Mode (PPM). Or else the converter emits a StreamElement only if the value read onto the OPC sever has changed (without the timestamp) between two updates - this is the Change Based Production Mode (CBPM).

We distinguish and allow two PPM cases and one CBPM case (see figure 2):

i) No OPC item is lost but it could be difficult to know the value of $\Delta t_S$ and to hold the synchronization.

ii) The converter produces more SreamElements than required and there is redundant information.

iii) In this case $\Delta t_P$ is not defined. Depending on the relationship between $\Delta t_S$ and $\Delta t_U$, the converter may miss some modifications of the OPC item values. However it cannot obviously generate redundant StreamElements as in case 2. The figure corresponds to $\Delta t_S > \Delta t_U$ and exactly each OPC item value change generates one StreamElement.

Clearly, we should have $\Delta t_{S,i} \geq \Delta t_U$ for *each* OPC item $i$ converted to be sure not to lose values from the OPC server. This requires the ability to change the production mode and $\Delta t_U$ during the execution.

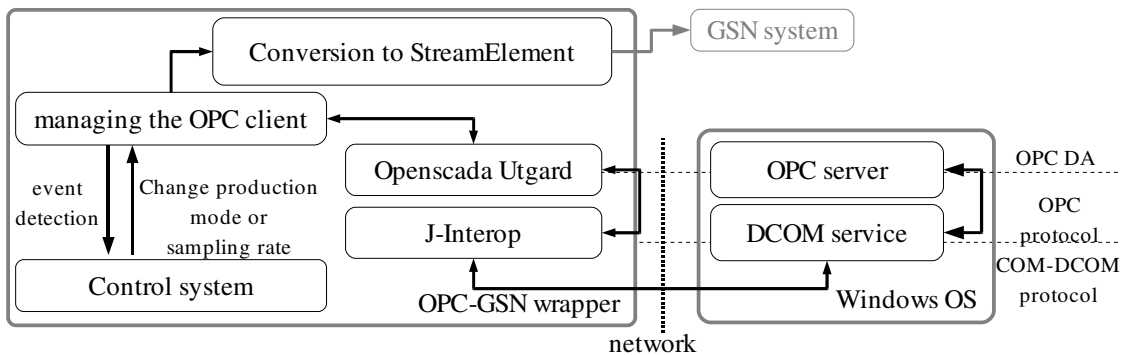

Figure 3: Software architecture of the OPC-GSN wrapper and its environment

## 4.2 Implementation architectures

The OPC-GSN converter has two possible positions in the system: i) Integrated converter: the converter is a software running on the Windows OS of the server (or a Windows OS COM-DCOM connected to the server). This solution requires the converter to generate data over HTTP connection with a GSN core, and to communicate with the OPC server through COM-DCOM services calls. ii)External converter: the OPC-GSN converter runs outside the Windows OS of the server and communicates with it through COM-DCOM: it can then be a GSN wrapper.

The last architecture is chosen for two reasons. First integrating OPC data inside a GSN system does not mean the ability to modify the running OPC system. It could then be forbidden or technically impossible to add a converter on the OPC side of the application. Second, removing the Windows OS constraint for the converter, we are able to provide an OPC-GSN connection running on any system supporting GSN and not only Windows OS.

### 4.3 OPC-GSN wrapper software

The wrapper conforms to OPC DA 2.05 and converts OPC data to GSN data type (see data types details in [AHS07a, BHMP07]). The wrapper is implemented in Java and supports two production modes (PPM and CBPM). Figure 3 gives the architecture of the wrapper. Connection to the OPC server is based on the Openscada Utgard project library [NSR06]. This project provides a Java OPC client and it uses the J-Interop library to manage the COM-DCOM protocol.

The upper left side of Figure 3 shows partially implemented features of the wrapper. Since we plan to use GSN as a Controlled and Adapted Distributed Information Fusion System, the wrapper will be controllable. In this respect, the current implementation allows the modification of the $\Delta t_U$ value and the kind of production mode at runtime.

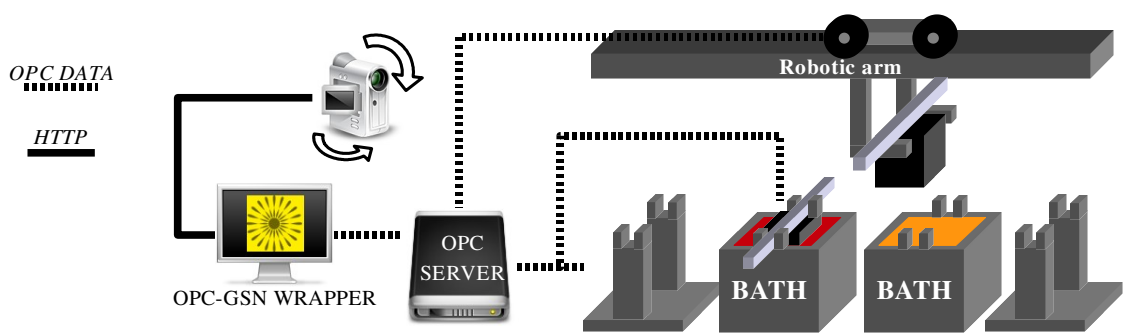

*Figure 4: Example of error detection: the GSN node monitors the presence of the work piece and a camera focuses if an error is detected.*

### 5. ERROR DETECTION APPLICATION

**The error detection.** This example is based on this scenario (see Figure 4): given a chain of surface treatment managed by an OPC server, and measurements processed by a GSN node connected with the OPC-GSN wrapper, we propose the following error detection. As soon as a work piece seems to be removed from a bath according to the sensor, the GSN node checks if the robotic arm is above the considered bath. If it is not, the error is risen and the camera focus on the bath. Then an operator can verify the presence of the work piece. In another case an automatic process can detect that the sensor of the bath is out of order and then removes it from the monitoring system and warns the user.

2 VS are used in this example. The first one contains the wrappers connected to the OPC server. We use several wrappers in order to allow different production modes at the same time on the same server. This VS also includes the alarm triggering requests that give as result the area where the error is detected. The second VS reads this result and sends the movement order to the camera. As it is easy to connect VSs together (only XML files have to be written), there is no limit in the creation of such information processing applications.

# 6. CONCLUSION

This article presents the design and implementation of a software component, termed wrapper in the GSN context, providing GSN compatible data from an OPC based system. This work fills the gap between industrial contexts and Information Fusion Systems (IFS): OPC servers are data sources consumed by the IFS built as a GSN infrastructure.

The OPC-GSN wrapper is restricted to the Data Access 2.05a specification of the OPC foundation to communicate with OPC servers. This is justified first by the large number of OPC servers still using this specification instead of the new OPC Unified Architecture and by the fusion process capabilities of GSN systems such as event detection and historical process.

We also decided to impact as less as possible OPC systems which will be connected to the GSN system. Our solution fits this requirement since there is no modification at all to do on the OPC side of the application using the OPC-GSN wrapper. All data processing is supported either by the wrapper or else by GSN features. A noticeable feature of OPC-GSN wrapper is the ability to set during the execution the refreshing period of data read from the OPC servers. This allows the IFS designer to adjust the semantics of the OPC data flow to its fusion process.

Our experiments in [BHMP07] show that the OPC-GSN wrapper accurately converts all data types of the OPC DA to GSN elementary information unit (StreamElement) and behave as expected with regard to the refreshing period.

This work is part of a global project on adaptable and controlled distributed information fusion systems. GSN infrastructures allows us to experiment various solutions with regard to the control of connections between fusion nodes and data sources and selection of data sources based on the goals of this information fusion system. In this respect, we will experiment control mechanisms for connections to data sources (OPC, sensor networks, etc). This will require extended control functionalities of the OPC-GSN wrapper.